\documentclass[aps,pre,twocolumn,groupedaddress,amsmath,amssymb]{revtex4-1}
\usepackage{graphicx}  
\usepackage{dcolumn}   
\usepackage{bm}        
\usepackage{verbatim}   
\usepackage{epstopdf}
\usepackage{subcaption}					
\usepackage{color}
\definecolor{orange}{rgb}{1,0.5,0}
\definecolor{brown}{rgb}{0.65, 0.16, 0.16}
\definecolor{phlox}{rgb}{0.87, 0.0, 1.0}
\graphicspath{{figs/}}					

\begin{document}
	
	\title{Modified Drude model for small gold nanoparticles surface plasmon resonance based on the role of classical confinement}
	\author {Asef Kheirandish}
	\affiliation{Department of Physics, University of Mohaghegh Ardabili, P.O. Box 179, Ardabil, Iran}
	\author {Nasser Sepehri Javan}
	\email{nsj108119@yahoo.com}
	\affiliation{Department of Physics, University of Mohaghegh Ardabili, P.O. Box 179, Ardabil, Iran}
	\author {Hosein Mohammadzadeh}
	\affiliation{Department of Physics, University of Mohaghegh Ardabili, P.O. Box 179, Ardabil, Iran}

	
		\begin{abstract}

		In this paper, we study the effect of restoration force caused by the limited size of a small metallic nanoparticle (MNP) on its linear response to the electric field of incident light. In a semi-classical phenomenological Drude-like model for small MNP, we consider restoration force caused by the displacement of conduction electrons with respect to the ionic positive background taking into account a free coefficient as a function of diameter of nanoparticle (NP) in the force term obtained by the idealistic Thomson model in order to adjust the classical approach. All important mechanisms of the energy dissipation such as electron-electron, electron-phonon and electron-NP surface scatterings and radiation are included in the model. In addition a correction term added to the damping factor of mentioned mechanisms in order to rectify the deficiencies of theoretical approaches. For determining the free parameters of model, the experimental data of extinction cross section of gold NPs with different sizes doped in the glass host medium are used and a good agreement between experimental data and results of our model is observed. It is shown that by decreasing the diameter of NP, the restoration force becomes larger and classical confinement effect becomes more dominant in the interaction. According to experimental data, the best fitted parameter for the coefficient of restoration force is a third order negative powers function of diameter. The fitted function for the correction damping factor is proportional to the inverse squared wavelength and third order power series of NP diameter. Based on the model parameters, the real and imaginary parts of permittivity for different sizes of gold NPs are presented and it is seen that the imaginary part is more sensitive to the diameter variations.  Increase in the NP diameter causes increase in the real part of permittivity (which is negative) and decrease in the imaginary part.
	\end{abstract}
	
	\maketitle
	
	
	\section{Introduction}
	Investigations dealing with the interaction of Electromagnetic Waves (EMWs) with metal Nanoparticles (NPs) and nanostructures are very attractive because of their fascinating applications in science and technology. The noble Metal Nanoparticles (MNPs) show a resonant interaction with EMWs in the visible spectrum which makes them ideal candidates for some special exotic applications in the industry and medicine. The origin of resonant interaction of MNPs is the collective oscillation of surface conduction free electrons with respect to the positive metallic lattice under interaction with light fields, i.e. surface plasmon. Occurrence of tremendous Electromagnetic (EM) fields enhancements in the resonance leads to the intense scattering and absorption of light \cite{mcconnell2000electronic,kreibig1995optical,faraday1857x} for MNPs. This plasmon resonance can either cause the light radiation via Mie scattering \cite{bohren2008absorption}, or the rapid conversion to heat through resonant absorption where both of mentioned processes play great roles in some new cutting-edge technological applications. Among the numerous applications of surface plasmon resonance of noble MNPs in new fields of science and technology, few important applications can be named as:  localized surface plasmon resonance sensing \cite{willets2007localized}, surface-enhanced Raman scattering spectroscopy \cite{nie1997probing}, surface-enhanced infrared absorption spectroscopy \cite{osawa2001surface,adato2009ultra}, enhanced nonlinear wave mixing \cite{renger2010surface,genevet2010large}, nano-scaled emission engineering, i.e. nano-antenna \cite{kosako2010directional}, optoelectronic devices \cite{maier2005plasmonics}, optical metamaterials \cite{pendry2003optics,pendry2006photonics}, solar cells \cite{atwater2011plasmonics}, frequency-sensitive photodetectors \cite{knight2011photodetection}, wavefront engineering of semiconductor lasers \cite{yu2010wavefront}, and molding light propagation at engineered interfaces \cite{yu2011light}. Beside the mentioned feasibilities of applications for noble MNPs related to their enhanced absorption and scattering, additionally, compositions of gold NPs are more suitable in medicine due to good biocompatibility, easy production \cite{burda2005chemistry} and ability to conjugate to a variety of biomolecular ligands and antibodies \cite{katz2004integrated} which make them very useful for NP-based cancer therapy \cite{hirsch2003nanoshell,huang2006cancer,el2006selective,o2004photo,loo2005immunotargeted,jain2007nanoparticles,huang2007gold,hirsch2003nanoshell2}, biological sensing \cite{elghanian1997selective,haes2002nanoscale}, imaging of bio-materials \cite{sokolov2003real,el2005surface,wang2004photoacoustic} and medical diagnostics \cite{rosi2005nanostructures}.\\
	Theoretical or experimental determination of complex dielectric permittivity or equivalently refractive index of media containing MNPs is necessary for explanation of dynamics of a lot of fundamental and applied phenomena and effects including the above mentioned subjects related to the absorption and scattering of light. In 1908, in the framework of classical electrodynamics by solving Maxwell’s equations, Mie \cite{bohren2008absorption} could obtain an analytical expression for extinction coefficient of a spherical particle describing the dissipation of light by absorption and scattering.  Up to date, Mie's calculations are the basis of the most of mathematical processes related to the interaction of NPs with the fields of EMWs. The crucial parameter needed for the extinction coefficient obtained by the Mie theory is the complex permittivity of MNPs which is calculated by the well-known phenomenological Drude theory. Theoretical considerations and experimental investigations have revealed that optical properties of NPs should dramatically depend on the size and shape of NPs. In several experimental studies, since 1958, Fragstein with his coworkers Roemer \cite{fragstein1958anomalie,romer1961bestimmung} and Schoenes \cite{von1967absorptionskoeffizient} have investigated the complex refractive index of silver NPs dispersed in colloidal solutions and determined the considerable differences between them and permittivity of bulk medium when the NPs dimensions were smaller than the mean free path of the conduction electrons. Then by Gans and Happel \cite{gans1909optik}, the effect of particles shape on the linear optical constants of such MNPs solutions has been calculated via Mie theory, so that one could determine the contribution of size and shape changes in the refractive index, separately. They measured linear optical constants of gold hydrosols and discovered that the changes in these values with respect to the bulk state are considerable only for the NPs including few atoms then they made conclusion that this difference arises from the additional electron scattering mechanism caused by the collision of conduction electrons with the particle surface that reduces their effective mean free path which is called the “free path effect”. To date, the free path effect is recognized as the main responsible factor for interpreting changes in the optical constants of MNPs with respect to the bulk state. There are a lot of theoretical and experimental studies where the modified Drude model is used for showing size dependence of MNPs permittivity only by adding free path effect on the damping coefficient of conduction electrons, just as a few examples see \cite{garcia2011surface,hovel1993width,voisin2001ultrafast,averitt1999linear,link1999size,kubo2007tunability,hartland2011optical,monteiro2012resonant,jain2006calculated,naik2013alternative,alvarez1997optical,pinchuk2004influence,grady2004influence,bruzzone2004light,alvarez1997optical,scaffardi2006size}.\\
	For an individual MNP, beside the occurrence of additional surface scattering caused by the limitation of the free path of conduction electrons, restriction of size of particle should lead to the appearance of restoring force caused by the displacement of electrons with respect to the background positive charges via exerting electric field of EMW. Such a linear restoring force reveals in atomic clusters using Thompson model \cite{kreibig1995optical}. Considering restoring force in dynamic equation of conduction electrons leads to the appearance of the resonance characteristic frequency of ${\omega _{res}} = {\omega _p}/\sqrt 3 $, where ${\omega _p}$ is the plasma frequency of conduction free electrons \cite{kreibig1995optical}. Sometimes ${\omega _{res}}$ is called the classical surface-plasmon frequency. Considering Mie theory for absorption of light by small MNPs with permittivity $\varepsilon$ doped in a background medium with permittivity ${\varepsilon _m}$, such a resonance frequency can be predicted when $\varepsilon  =  - 2{\varepsilon _m}$ \cite{bohren2008absorption}. In the simplifying limit of $\Gamma  <  < {\omega _p}$, where $\Gamma $ is the damping factor related to the various mechanisms of electron scattering, this condition leads to the maximum absorption of light at the resonance frequency ${\omega _{res}} = {\omega _p}/\sqrt {1 + 2{\varepsilon _m}} $ which reduces to ${\omega _p}/\sqrt 3 $ in the case of considering air as the surrounding medium, i.e. ${\varepsilon _m} = 1$ \cite{bohren2008absorption}. It would be interesting to know that in such resonance condition, the polarizability of an MNP exposed to an electric field shows resonant behavior as well \cite{kreibig1995optical}. All the mentioned facts in this paragraph, may emphasize on the importance of considering restoring force in dynamical problems related to the interaction of EMWs with MNPs in the classical regime. In some studies related to the interaction of EMWs with systems including MNPs, in classical momentum equation, the restoring force which can predict the resonance frequency is considered \cite{sepehri2015nonlinear,sepehri2015raman,sepehri2015self,sepehri2016dielectric,sepehri2017magnetic,sepehri2017theoretical,kheirandish2018polarization,sepehri2018theoretical,javan2019magnetic,javan2019semi,hernandez2005asymmetric,parashar2009stimulated,kumar2007anomalous,brongersma2000electromagnetic} but surprisingly it is absent in the dynamics of the most of the studies related to the Drude-based problems and as it is mentioned in the previous paragraph, they considered only the free path limitation on the dynamics of conduction electrons in MNPs. In some studies related to the plasmonics of metallic nanostructures, for extraction of permittivity of system by Drude-based models, in the equation of motion of damped harmonic oscillation, the restoring force is considered and the related spring constant is determined by simulation \cite{zuloaga2011energy,kats2011effect,biagioni2012nanoantennas,lovera2013mechanisms,yu2014flat,genevet2017recent}. In this article using a simple Drude-like model that considers the restoration force, the complex permittivity of an individual gold NP is studied. Important mechanisms of conduction electron scattering including electron-electron, electron-phonon, electron-surface and radiation as well are considered via well-known theoretical relations and a correction term is added in order to correct some shortages of theories. Also, a correction coefficient is considered for the restoring force in order to rectify the shortages of ideal model of Thompson. Estimation for free parameters of system is accomplished by considering experimental data for extinction coefficient of gold NPs with different sizes doped in the glass. Results show a good agreement between experiments and our model.
\section{Thomson model for small MNP}
We use classical Thomson model \cite{thomson1904xxiv} for describing interaction of EMW with spherical MNP. Even though this model was unsuccessful for describing atomic structure but it is still convenient for constructing classical theories in the light-cluster interaction phenomena [2]. In this model, it is supposed that conduction electrons of $N$ individual atoms are homogenously distributed inside a sphere with radius $R$ and background positively charged ions which are distributed homogenously as well, are immobile. If the average separation of atoms is d, then the density of atoms or equivalently the density of background ions is ${n_a} = 1/{d^3}$ while we denote the density of conduction electrons as ${n_e} = ZNe/[(4/3)\pi {R^3}]$, where $e$ is the magnitude of electron charge and $Z$ is the number of conduction electrons for each atom. For a small NP exposed to the low-intensity EM fields where its radius is much less than the wavelength, ${\kern 1pt} R <  < \lambda $, one can neglect the spatial variation of EM fields inside the NP and suppose that the same forces act on all conduction electrons at a moment. The motion equation of electrons inside the NP confined to the radius $R$, can be written as 
\begin{align}
&m{{\bf{\ddot x}}_{\bf{i}}} =  - e{{\bf{E}}_{\bf{0}}}{\bf{(}}{{\bf{x}}_{\bf{i}}},t{\bf{)}} - e{{\bf{E}}_{\bf{p}}}{\bf{(}}{{\bf{x}}_{\bf{i}}},t{\bf{)}} - m\Gamma {{\bf{\dot x}}_{\bf{i}}} \nonumber \\ 
&+ \frac{1}{{4\pi {\varepsilon _0}}}\sum\limits_{j \ne i}^{ZN} {\frac{{{e^2}}}{{{{\left| {{{\bf{x}}_{\bf{i}}} - {{\bf{x}}_{\bf{j}}}} \right|}^3}}}} \left( {{{\bf{x}}_{\bf{i}}} - {{\bf{x}}_{\bf{j}}}} \right),\,\,\,\,i=1,2,...,ZN,
\end{align}
where $m,\,{{\bf{x}}_{\bf{i}}},\,{\varepsilon _0}$ and $\Gamma$ are electron mass, the ith electron position vector, permittivity of vacuum, and damping factor related to any kind of energy dissipation mechanisms, respectively. At the right side of Eq. (1), the first term is the force of external electric field, the second term describes the force caused by the background positive ions and the last term denotes the electron-electron interaction. Using Gauss's law, one can easily obtain the following equation for the electric field of the background ions
\begin{equation}
{{\bf{E}}_{\bf{p}}} = \frac{{e{n_e}}}{{3{\varepsilon _0}}}{{\bf{x}}_{\bf{i}}} = \frac{{m{\omega _p}^2}}{{3e}}{{\bf{x}}_{\bf{i}}},
\end{equation}
where ${\omega _p} = \sqrt {{n_e}{e^2}/m{\varepsilon _0}} $ is the plasma frequency. Introducing the well-known concept of center of mass for conduction electrons as
\begin{align}
	{\bf{X}} = \frac{1}{{ZN}}\sum\limits_{i = 1}^{ZN} {{{\bf{x}}_{\bf{i}}}},
\end{align}
and using it in Eq. (1) after doing summation on motion equations of the whole electrons existing inside the NP, one can reach to the following well-known damped harmonic oscillation equation for the center of mass displacement
\begin{align}
	{\bf{\ddot X}} + \Gamma {\bf{\dot X}} + \frac{{{\omega _p}^2}}{3}{\bf{X}} =  - \frac{e}{m}{{\bf{E}}_{\bf{0}}}{\bf{(t)}},
\end{align}
where the term related to the electron-electron interactions vanishes during summation process because of the action-reaction law.
\section{Modified Drude model}
For a metal bulk system, considering free electrons via ignoring the third term in the left side of Eq. (4) and taking into account a monochromatic field oscillating with frequency $\omega $, i.e. ${E_0} \sim {e^{ - i\omega t}}$, one can obtain the following equation for the permittivity of bulk metal
\begin{align}
	{\varepsilon _{bulk}} = 1 - \frac{{{\omega _p}^2}}{{{\omega ^2} + i{\Gamma _{bulk}}\omega }},
\end{align}
where we used index “bulk” to distinguish bulk system with confined one. Considering the role of inner electrons in atoms, Eq. (5) reads
\begin{align}
{\varepsilon _{bulk}} = {\varepsilon _\infty } - \frac{{{\omega _p}^2}}{{{\omega ^2} + i{\Gamma _{bulk}}\omega }},
\end{align}
where ${\varepsilon _\infty }$ reflects interaction of inner electrons with light and itself can be written in the form \cite{bohren2008absorption}
\begin{equation}
{\varepsilon _\infty } = 1 + \sum\limits_{j = 1}^{{N_o}} {\frac{{{\omega _{pj}}^2}}{{({\omega _j}^2 - {\omega ^2}) - i{\Gamma _j}\omega }}},
\end{equation}
where ${N_o}$ denotes the number of Lorentz oscillators, j presents the special kinds of electrons located at inner levels having similar behavior during interaction with light fields, ${\omega _{pj}},\,{\omega _j}$ and ${\Gamma _j}$ are the plasma frequency related to the special kind of electrons population, their resonant frequency and their damping factor, respectively, which can be measured experimentally.\\
For an NP with limited size, in Eq. (4), considering the third term related to the restoration force leads to the special resonance frequency for free electrons at ${\omega _p}/\sqrt 3 $ which called plasmon frequency. As experimental measurements show, the place of plasmon frequency extremely differs from ${\omega _p}/\sqrt 3 $ which can be referred to the existence of electron damping. As mentioned in the introduction section, in the most of works related to the calculation of optical parameters of NPs, restoring force is ignored from dynamical models and the size effect is only considered in damping factor by introducing new so-called surface scattering mechanism which is caused by the limitation of mean free path of electrons confined in an NP. Even though, for NP whose radius is greater than or comparable with light wavelength, considering the idealistic model in which all conduction electrons treat in the same manner, cannot be correct, however effect of background ions on electrons which reflects the classical confinement characteristic of system cannot be ignored. Here, we consider the restoration force by multiplying it with a coefficient which is a function of radius and introduce the permittivity of an individual NP as 
\begin{align}
	{\varepsilon _n} = {\varepsilon _{bulk}} + \frac{{{\omega _p}^2}}{{{\omega ^2} + i{\Gamma _{bulk}}\omega }} - \frac{{{\omega _p}^2}}{{{\omega ^2} - \alpha {\omega _p}^2 + i{\Gamma _n}\omega }},
\end{align}
where the sum of two first terms of the right hand is ${\varepsilon _\infty }$ for the bulk metal, $\Gamma_{Bulk}=0.07 \times 10^{15}s^{-1}\,\,\,$ \cite{myroshnychenko2008modelling} and the last term denotes the contribution of conduction electrons of limited NP. $\alpha $ is a function of NP radius which should vanish for large particles and in ideal case of zero radius should limit to the well-known value of $1/3$. ${\Gamma _n}$ stands for the damping factor of electrons in a confined region of NP and will be calculated as
\begin{align}
	{\Gamma _n} = {\Gamma _{e - e}} + {\Gamma _{e - ph}} + {\Gamma _{rad}} + {\Gamma _{surf}} + {\Gamma _{cor}},
\end{align}
  ${\Gamma _{e - e}}$ is the damping factor related to the scattering of an electron by another one in a bulk lattice of metal which can be calculated by the well-known theoretical relationship derived by Lawrence and Wilkins \cite{lawrence1976electron,lawrence1973electron}
\begin{align}
	{\Gamma _{e - e}} = \frac{{{\pi ^2}}}{{24\hbar {E_F}}}\left[ {{{({k_B}T)}^2} + {{(\hbar \omega )}^2}} \right],
\end{align}
where $\hbar  = h/(2\pi ),\,h$ is the Planck’s constant, ${E_F}$, ${k_B}$ and $T$  are the Fermi energy, the Boltzmann’s constant and temperature, respectively.\\
  ${\Gamma _{e - ph}}$ is the damping factor concerned with the energy dissipation due to the interaction of conduction electrons with metallic lattice which is obtained using the theoretical relation derived by Holstein \cite{holstein1954optical,holstein1964theory} as following
\begin{align}
	{\Gamma _{e - ph}} = {\Gamma _0}\left( {\frac{2}{5} + \frac{{4{T^5}}}{{\theta _D^5}}\int_0^{\frac{{{\theta _D}}}{T}} {\frac{{{z^4}}}{{{e^z} - 1}}dz} } \right),
\end{align}
where ${\theta _D}$ is the Debye temperature and ${\Gamma _0}$ is a constant that can be achieved through the fitting procedure of the bulk permittivity for the frequency interval which is located below the interband transition threshold.\\
  ${\Gamma _{rad}}$ denotes the damping factor caused by the radiation of accelerated electrons and it can be derived by classical electrodynamic considerations using Abraham-Lorentz force as the following simple relation \cite{liu2009reduced}
\begin{align}
	{\Gamma _{rad}} = \frac{{\omega _p^2V{\omega ^2}}}{{6\pi {c^3}}},
\end{align}
where  $V=(4/3)\pi R^3$ is the NP volume and c is the light speed in vacuum. \\
The damping factor ${\Gamma _{surf}}$ is related to the limitation of mean free path of electrons bounded inside an MNP. It can be calculated through the scattering mechanism of a free electron by the surface of MNP. We use the following formula for considering contribution of size limitation in the energy dissipation \cite{genzel1975dielectric}
\begin{align}
	{\Gamma _{surf}} = \frac{{A{v_f}}}{{{L_{eff}}}}.
\end{align}
  where $A$ is a dimensionless parameter whose value can be obtained considering some details of the scattering mechanism and has own scientific story which will be mentioned briefly, ${v_f} = 1.4 \times {10^6}m/s$ is the Fermi velocity for gold and ${L_{eff}}$ is the reduced mean free path of electrons. Employing a geometrical probability method, Coronado and Schatz \cite{coronado2003surface} extracted the following simple equation for the reduced mean free path ${L_{eff}} ={{4V}}/{S}$, where  $V$ and $S$ are the volume and surface area of an NP with arbitrary curved shape. This term is related to totally inelastic scattering of conduction electrons by the surface of MNP and states the mean chord distance of any two arbitrary points located on the particle surface\cite{liu2004synthesis}. Even though we know that parameter $A$ should empirically have relation with the shape of MNP, however determining its value is the place of argument. In Ref. \cite{coronado2003surface}, a value near to unity is suggested to the parameter $A$, but value $A = 0.33$ is proposed by Berciaud et al.\cite{berciaud2005observation} by fitting values of parameters related to the absorption of light by the single gold NP.\\
  ${\Gamma _{cor}}$ is a correction term which we add to the damping factor in order to adjust theoretical formulae by considering experimental data.  Two first terms of Eq. (9) are related to the dominant dissipation mechanisms for the bulk metal system which should be corrected for the limited size of an NP because of the quantum confinement effects. On the other hand some mechanisms like existence of defects in the lattice does not take into account. Therefore, in order to match theoretical results with experimental ones we will determine this term by trial and error method which will be explained in numerical section in detail.
\section{Numerical Analysis}
To extract the free parameters of the model, i.e. $\alpha $ and ${\Gamma _{cor}}$, we use some experimental data related to the linear interaction of light with NPs. Extinction cross section of media including MNPs is the well-known experimental data that we can employ them to guess the model parameters. The extinction cross section is defined as the sum of the absorption cross section and the scattering one which can be expressed as following by using Mie theory \cite{bohren2008absorption}
\begin{figure}[t]
	\centerline{\includegraphics[scale=.25]{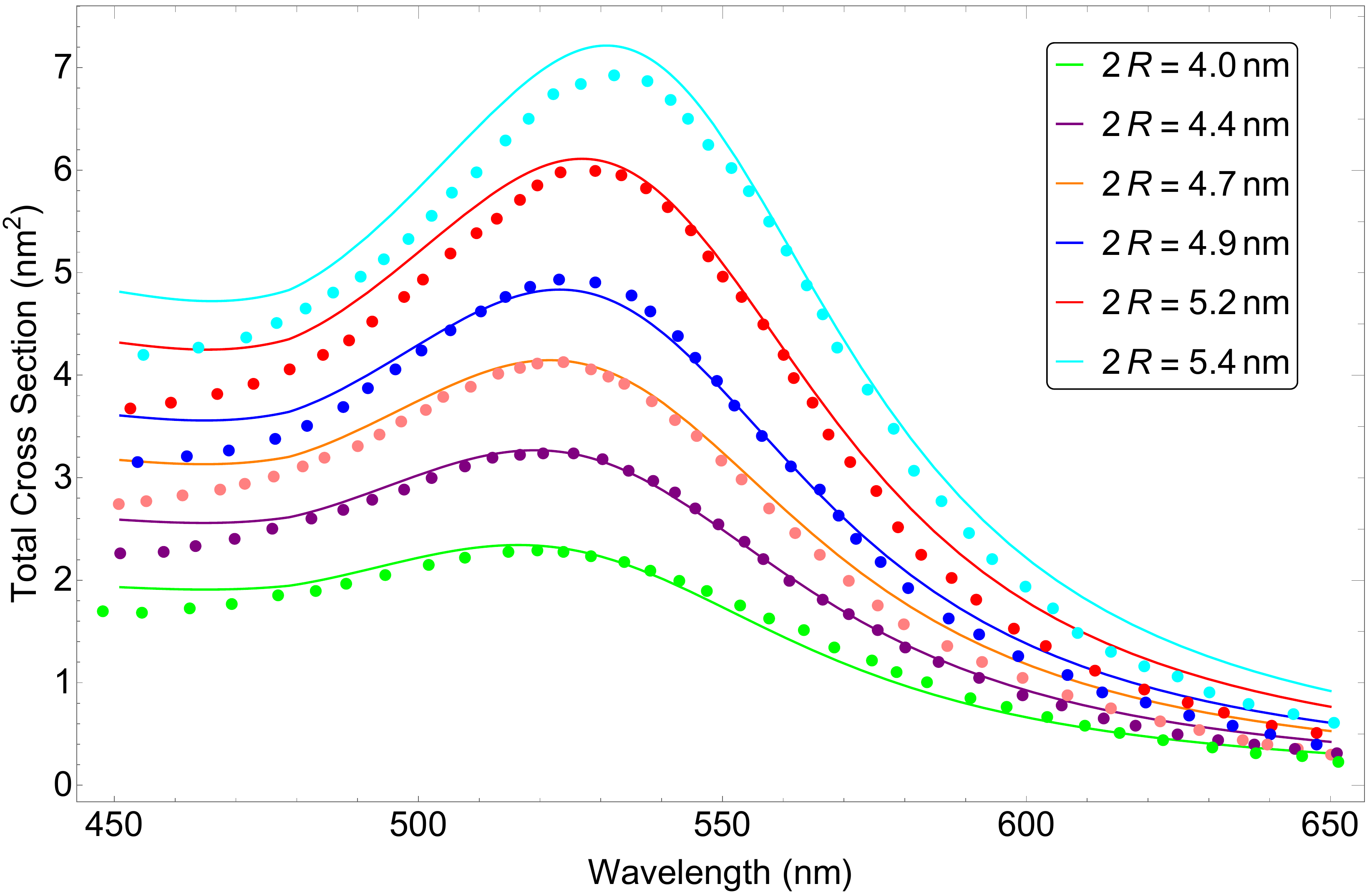}}
	\caption{The calculated extinction cross section (solid line) and the experimentally measured one (dotted line) in dependence of wavelength for NP diameters of 4, 4.4, 4.7, 4.9, 5.2, 5.4nm (from bottom to top).}
	\label{fig1}
\end{figure}
\begin{align}
{C_{ext}} &= \frac{{2\pi }}{{{k^2}}}\sum\limits_{n = 1}^\infty  {(2n + 1){\mathop{\rm Re}\nolimits} ({a_n} + {b_n})} ,\\
{a_n} &= \frac{{m{\psi _n}(mx){{\psi '}_n}(x) - {\psi _n}(x){{\psi '}_n}(mx)}}{{m{\psi _n}(mx){{\xi '}_n}(x) - {\xi _n}(x){{\psi '}_n}(mx)}},\\
{b_n} &= \frac{{{\psi _n}(mx){{\psi '}_n}(x) - m{\psi _n}(x){{\psi '}_n}(mx)}}{{{\psi _n}(mx){{\xi '}_n}(x) - m{\xi _n}(x){{\psi '}_n}(mx)}},
\end{align}
where $x=2\pi NR/\lambda$ is the size parameter, $m=N_1/N$ is the relative refractive index, where $N_1$ and $N$ are the refractive indices of particle and background medium, respectively, ${\psi _n}(x)$ and ${\xi _n}(x)$ are Riccati-Bessel functions.\\
For small particles, i.e. $R/\lambda  \ll 1$, the extinction cross section reduces to
\begin{figure}[t]
	\centerline{\includegraphics[scale=.3]{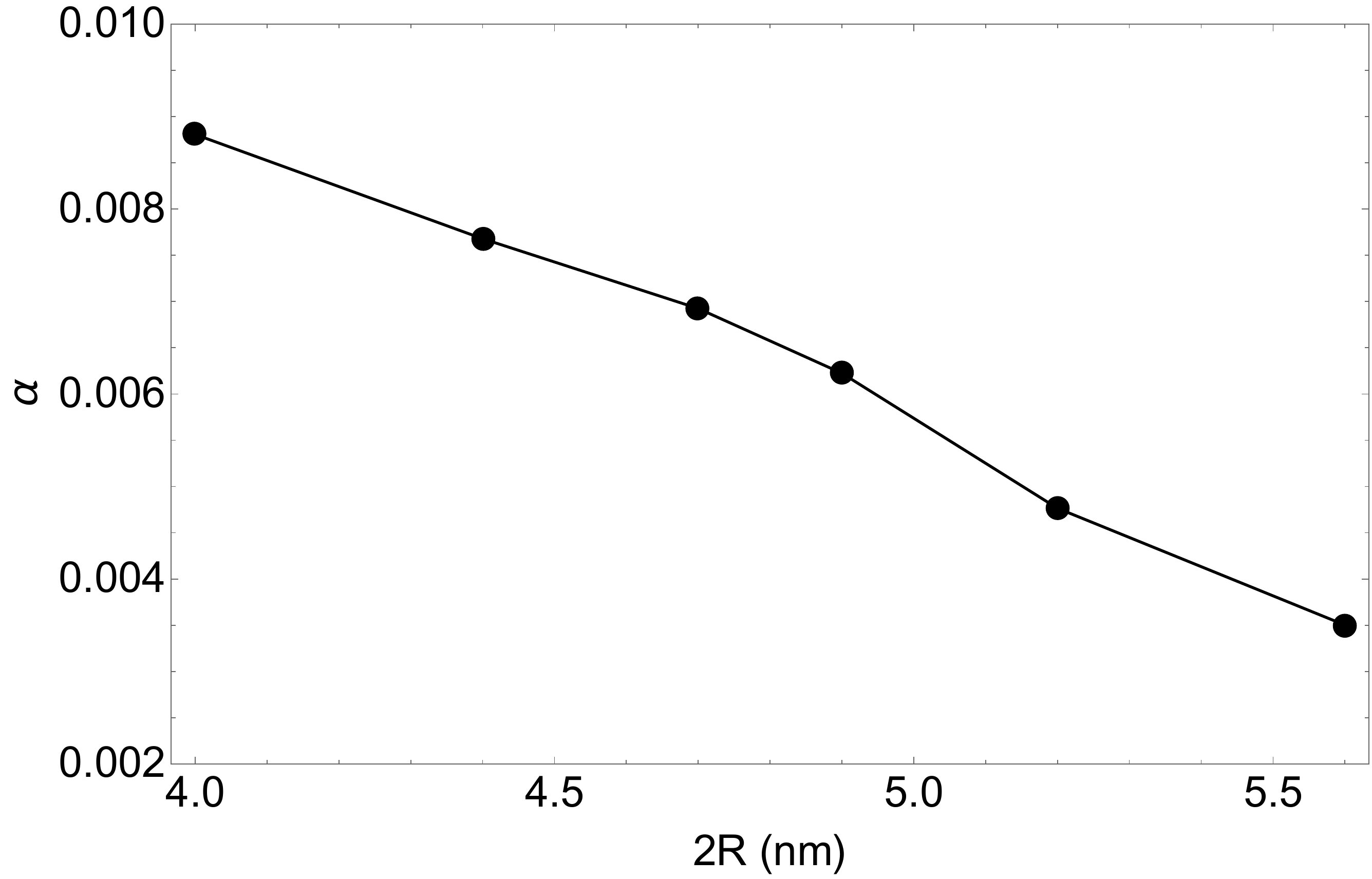}}
	\caption{Variations of parameter $\alpha$ as a function of NP diameter}
	\label{fig2}
\end{figure}
\begin{align}
{C_{ext}} &= 4x\pi {R^2}{\mathop{\rm Im}\nolimits} \left\{ {\frac{{{m^2} - 1}}{{{m^2} + 2}}\left[ {1 + \frac{{{x^2}}}{{15}}\left( {\frac{{{m^2} - 1}}{{{m^2} + 2}}} \right)} \right.} \right. \nonumber \\
\times & \left. {\left. {\frac{{{m^4} + 27{m^2} + 38}}{{2{m^2} + 3}}} \right]} \right\} + \frac{8}{3}{x^4}{\mathop{\rm Re}\nolimits} \left[ {{{\left( {\frac{{{m^2} - 1}}{{{m^2} + 2}}} \right)}^2}} \right]
\end{align}
which in the case of very small particles or linear regime of $x$, limits to the well-known relationship for Rayleigh scattering as following  
\begin{equation}
{C_{ext}} = 4x\pi {R^2}{\mathop{\rm Im}\nolimits} \left( {\frac{{\varepsilon  - {\varepsilon _m}}}{{\varepsilon  + 2{\varepsilon _m}}}} \right).
\end{equation}
In Fig. 1, the extinction cross section of an individual gold NP doped in a glass background medium has been plotted for different sizes of small spherical NPs $(2R<10nm)$. The dotted lines are obtained experimentally by Kreibig and Vollmer \cite{kreibig1995optical} and the solid ones show our model results. Trial and error procedure for obtaining best fit for the extinction cross section reveals that the best fitted functions for $\alpha $ and ${\Gamma _{cor}}$ are as following
\begin{align}
&\alpha  = {\alpha_0} + {c_1}\frac{1}{R} + {c_2}\frac{1}{{{R^2}}} + {c_3}\frac{1}{{{R^3}}},\\
&{\Gamma _{cor}} \times 10^{-20} = {d_1}\frac{1}{{{\lambda ^2}}} + {d_2}\frac{{{R^2}}}{{{\lambda ^2}}} + {d_3}\frac{{{R^3}}}{{{\lambda ^2}}},
\end{align}
where
\begin{align}
{{\rm{\alpha}}_0}&{\rm{ =  - 0}}{\rm{.314,}}\,\,{{\rm{c}}_1}{\rm{ = 2}}{\rm{.052nm,}}\,\,\,{{\rm{c}}_2}{\rm{ =  - 4}}{\rm{.399nm^2,}}\nonumber \\
{{\rm{c}}_3}&{\rm{ = 3}}{\rm{.176nm^3,}} \\
{d_1} &= -1.084nm^2,{\kern 1pt} \,\,{d_2} =  8.615 \times {10^{ - 2}},\,\,{\kern 1pt}\nonumber \\
{d_3} &= -3.883 \times {10^{ - 2}nm^{-1}},
\end{align}
and all lengths are in nm.\\
It should be better to mention that we could choose other functionalities for free parameters in order to exactly fit the model and experimental data, however we choose the above forms because of their simplicity and being physically meaningful as well. The first and third terms of ${\Gamma _{cor}}$ have the same form of the electron-electron and radiation scattering terms with respect to the radius of NP and light wavelength. These terms are negative and it indicates that the total amounts of the mentioned scattering terms become smaller by considering experimental corrections. The second term in Eq. (20) which is positive, has the functionality form of ${(R/\lambda )^2}$ and it can be interpreted as the contribution of other ignored mechanisms of scattering and quantum corrections as well. The parameter $\alpha $ is the function of negative powers of R and it is independent of wavelength. In Fig. (2), the parameter $\alpha $ is presented as a function of NP diameter in nanometer. Increase in the NP diameter causes the decrease in $\alpha $ which is a logical behavior. Its value decreases approximately from $0.009$ to $0.0035$ when NP diameter increases from $4nm$ to $5.4nm$, respectively. As it is mentioned earlier, we expect that by growing the size of NP, classical confinement effect (or in other words, appearing the restoring force) fades out and for large size NPs, it vanishes. The largest value for $\alpha $ which can be predicted by primitive classical theories is $1/3$.\\
\begin{figure}[b]
	\begin{subfigure}{0.40\textwidth}\includegraphics[width=\textwidth]{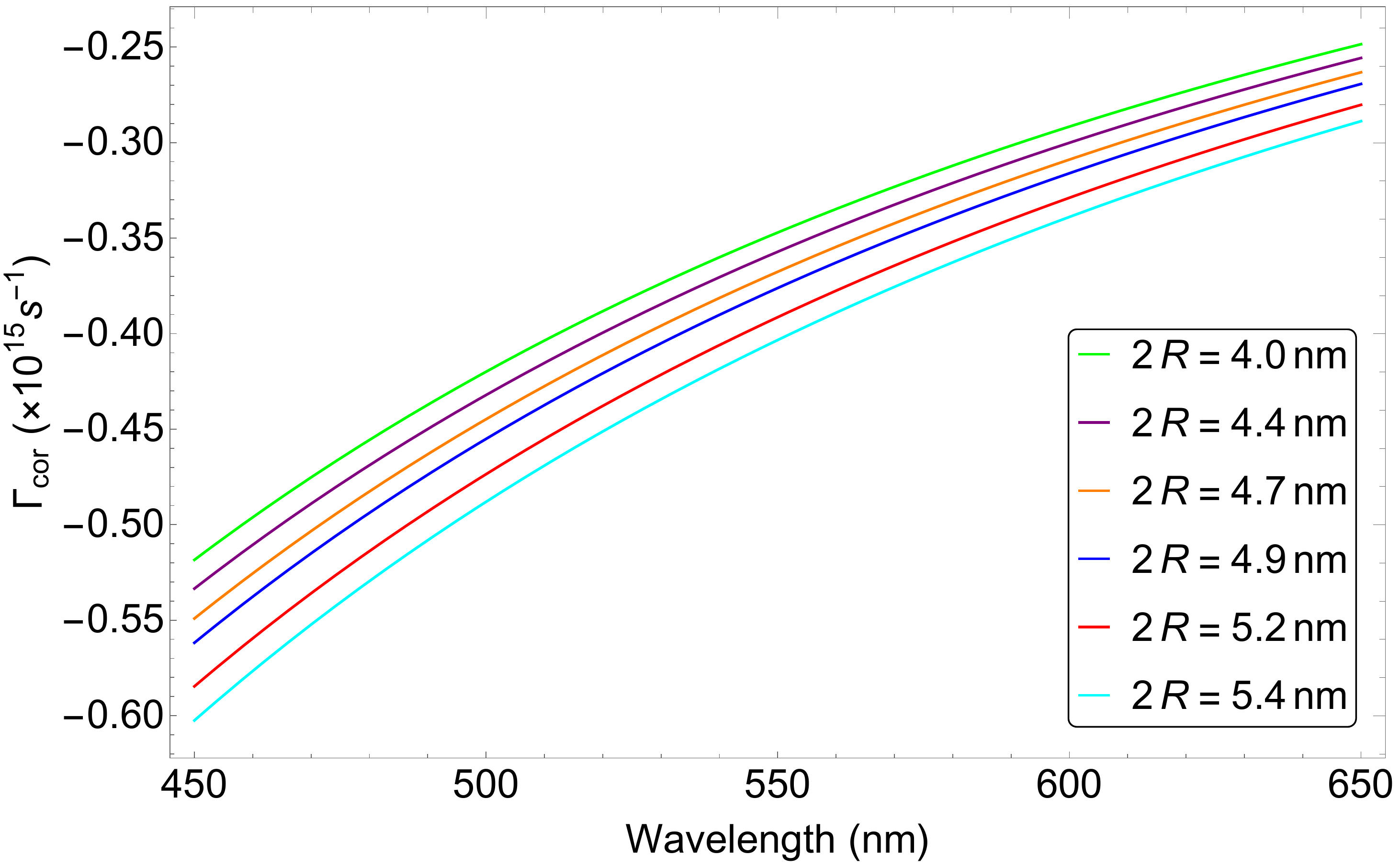}
		\caption{}
		\label{fig3a}
	\end{subfigure}
	\begin{subfigure}{0.40\textwidth}\includegraphics[width=\textwidth]{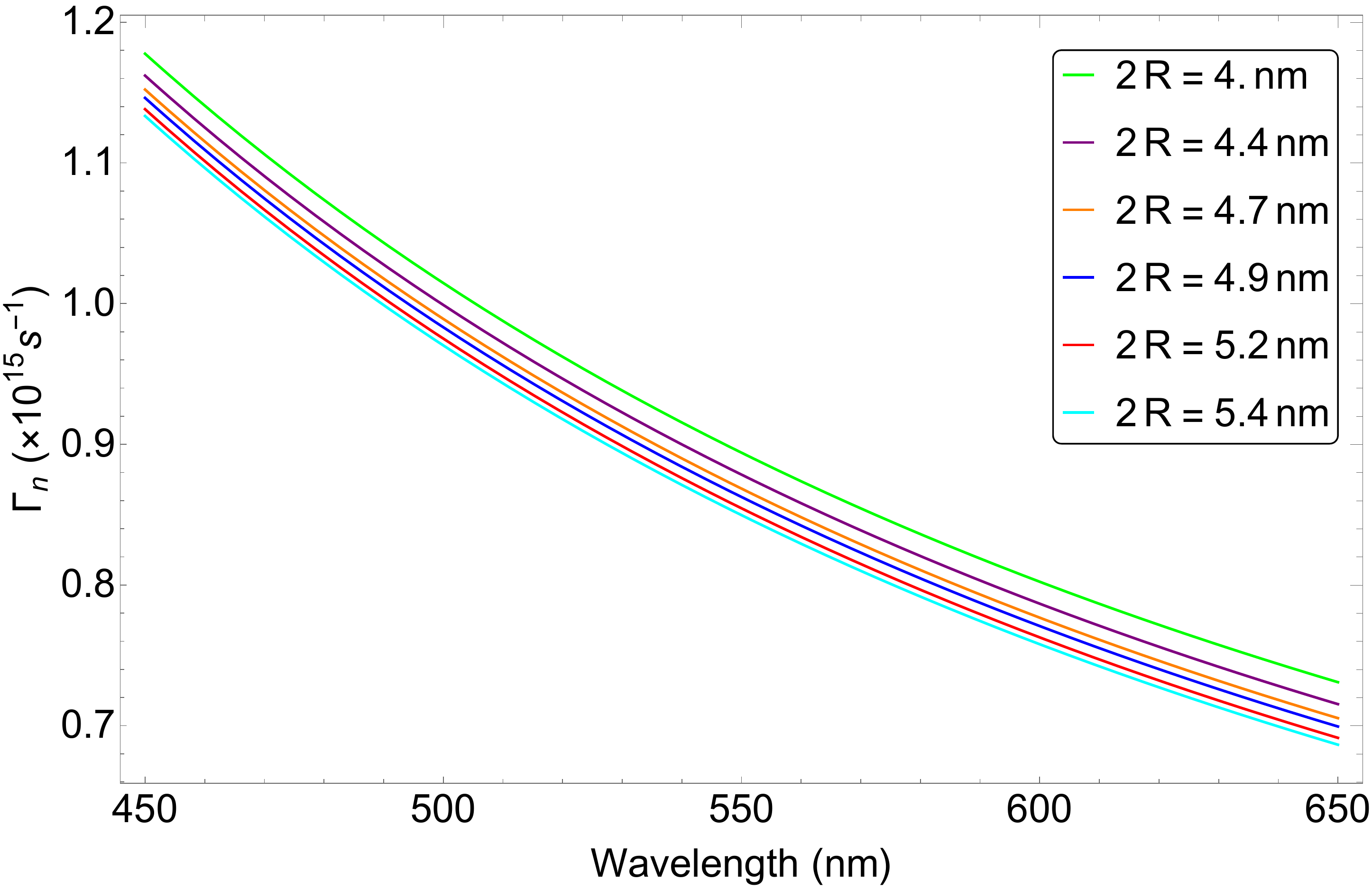}
		\caption{}
		\label{fig3b}
	\end{subfigure}
	\caption{Variations of (a)	$\Gamma_{Cor}$ and (b) $\Gamma_{n}$ as a function of wavelength for nanosphere diameters of 4, 4.4, 4.7, 4.9, 5.2, 5.4nm. The order of diameter increase is from top to bottom.}
	\label{fig:fig5}
\end{figure}
In Fig. (3-a) and Fig. (3-b), we have plotted $\Gamma _{cor}$ and $\Gamma _{n}$ as a function of light wavelength for different NP sizes. For all cases, $\Gamma _{cor}$ is negative and its absolute value is an increasing function of NP diameter which in turn causes that the parameter $\Gamma _{n}$ becomes a decreasing function of NP size. It is clear that increase in the wavelength causes the decrease in both parameters $\left| {{\Gamma _{cor}}} \right|$ and $\Gamma _{n}$ at a fixed NP diameter size. The average value of $\Gamma _{n}$ for small size NPs varies from $10^{15}s^{-1}$ to $0.7 \times 10^{15}s^{-1}$ when wavelength changes from $450nm$ to $650nm$, respectively.\\
\begin{figure}
	\begin{subfigure}{0.42\textwidth}\includegraphics[width=\textwidth]{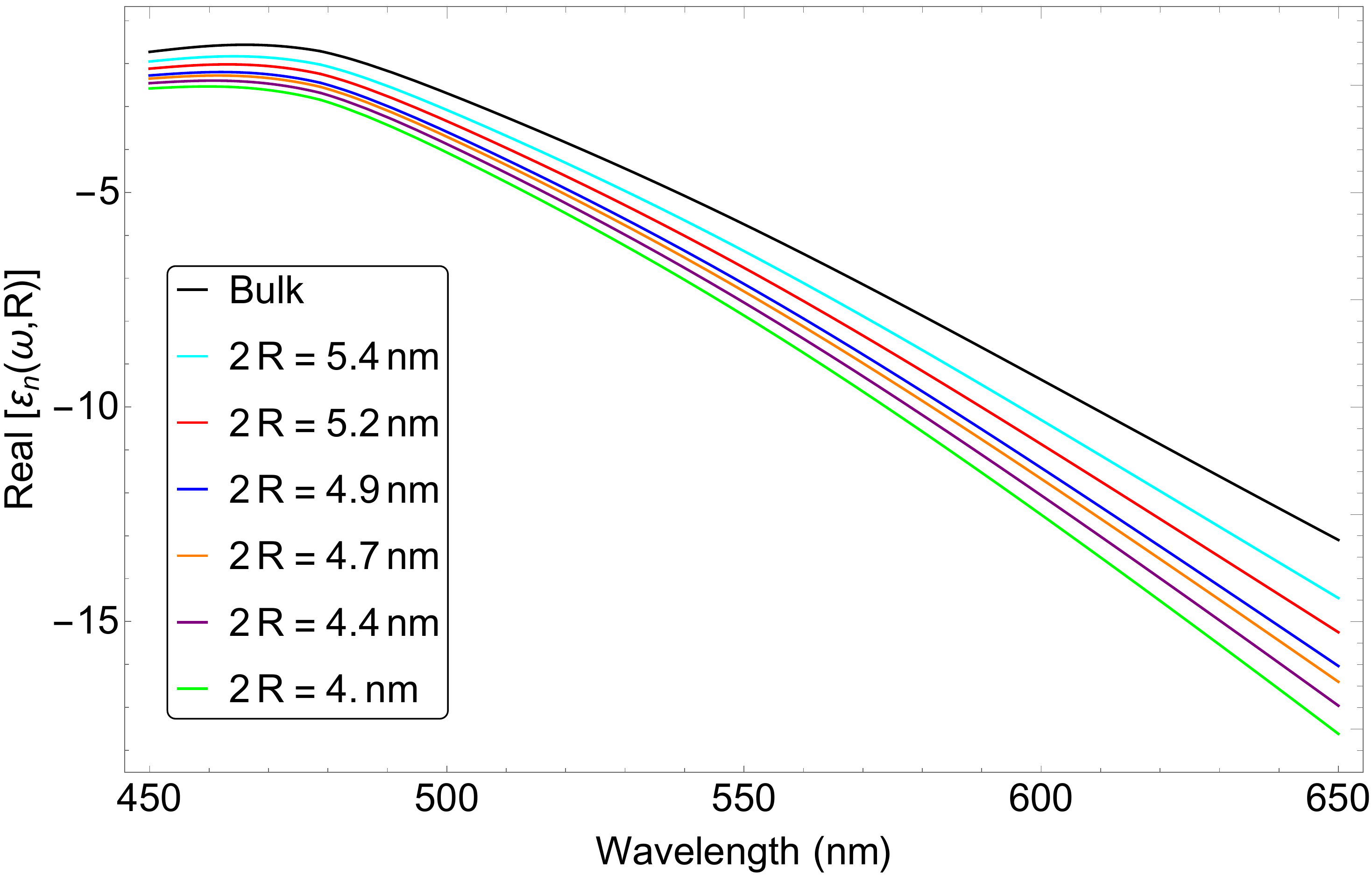}
		\caption{}
		\label{fig5a}
	\end{subfigure}
	\begin{subfigure}{0.40\textwidth}\includegraphics[width=\textwidth]{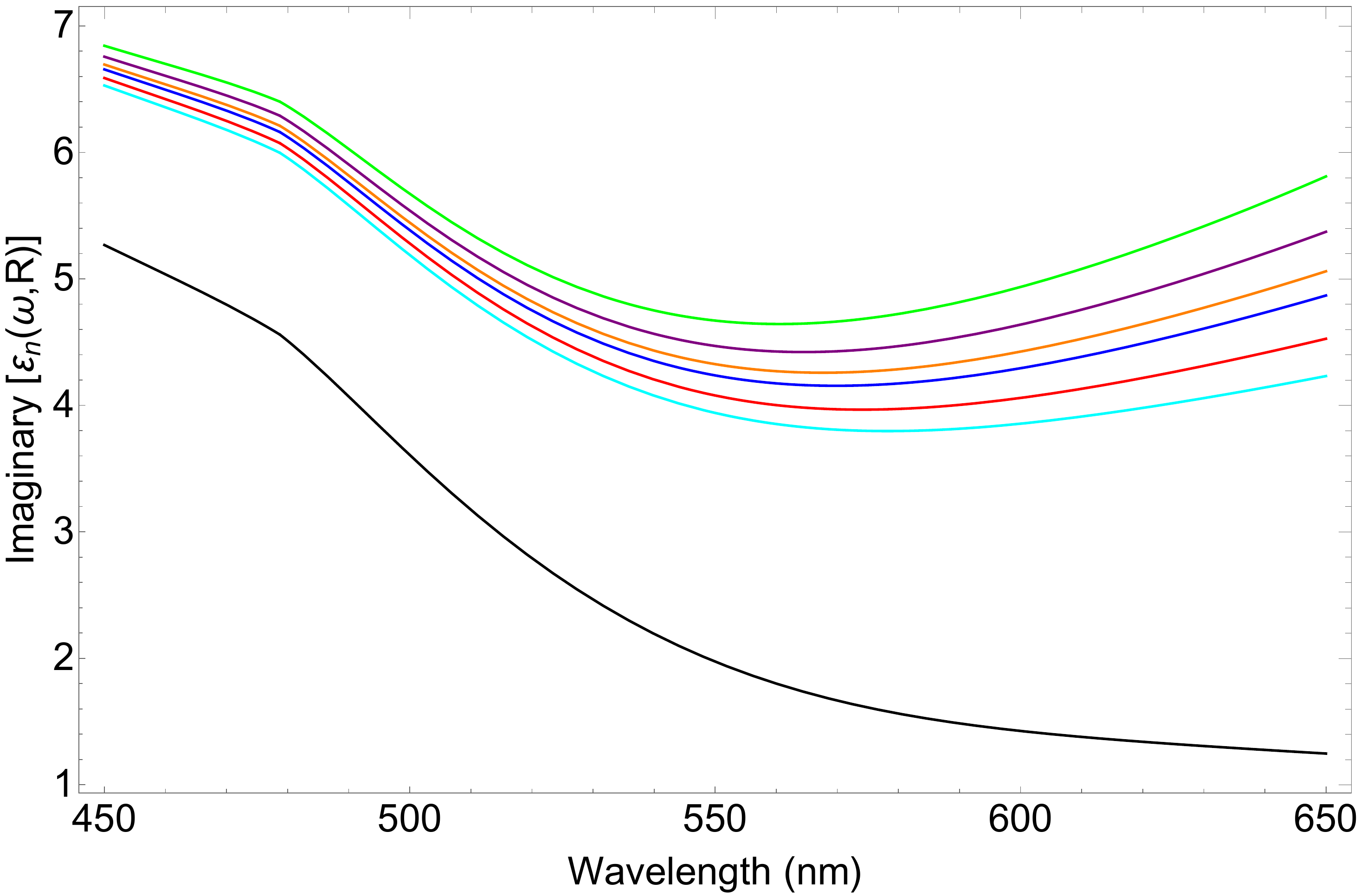}
		\caption{}
		\label{fig5b}
	\end{subfigure}
	\caption{Variations of (a) real and (b) imaginary parts of permittivity as a function of wavelength for nanosphere diameters of 4, 4.4, 4.7, 4.9, 5.2, 5.4nm and bulk case as well. The order of diameter increase for graphs (a) is from bottom to top and for graphs (b) is inversely.}
	\label{fig:fig6}
\end{figure}
In Fig. (4-a) and Fig. (4-b), the real and imaginary parts of gold NP permittivity is presented for different small size NPs and bulk gold metal as well. The data of bulk medium have been taken from Ref \cite{johnson1972optical}. As an example for metal, the real part of permittivity is negative for all cases and in a fixed wavelength, increase in the diameter of NP causes the decrease in the absolute value of real part. It is clear from Fig. (4-b) that increase in the size of NP causes the decrease in the imaginary part of permittivity. Totally, variations of permittivity values with respect to the NP size variations are more considerable for the imaginary part which reflects the absorption characteristic of medium. Especially, dependence of imaginary part on the NP size is more evident at high wavelengths or low photon energies. There are no experimental data for direct measurement of gold NP permittivity and only the permittivity of gold thin films can be found in Refs \cite{johnson1972optical,hagemann1975optical,ordal1987optical,windt1988optical,palik1998handbook,werner2009optical,mcpeak2015plasmonic,babar2015optical}. Recently, Karimi et al.\cite{karimi2018surface} have been proposed a size-dependent plasma frequency model for MNP permittivity in quantum regime and similar theoretical results are obtained for the real and imaginary parts of gold small NPs. They interpreted the intense dependence of imaginary part on the NP size at low photon energies as the result of the intense increase in the surface scattering of NP.\\
In order to show the size-dependence of plasmon resonance predicted by our semi-classical phenomenological model and compare them with experiments, in Fig. (5), we have plotted the plasmonic peak wavelength (or surface plasmon resonance peak wavelength) as a function of NP diameters. Increase in the NP diameter increases the resonance wavelength, or in other words causes a redshift in resonance wavelength. One can see the good agreement between our model and experimental data. The dependence of peak place to the NP diameter can be expressed via the following function
\begin{align}
	{\lambda _{\max }} = {\lambda_0} + {f_1}R + {f_2}{R^2} + {f_3}{R^3},
\end{align}
where
\begin{align}
	&{\lambda_0} = 303.90nm,\,\,{f_1} = 303.20nm^{-1},\,\,{f_2} =  - 146.65nm^{-2},\nonumber \\
	&{f_3} = 24.28nm^{-3},
\end{align}
where 	${\lambda _{\max }}$ and $R$ are in nm.
\begin{figure}[h]
	\centerline{\includegraphics[scale=.3]{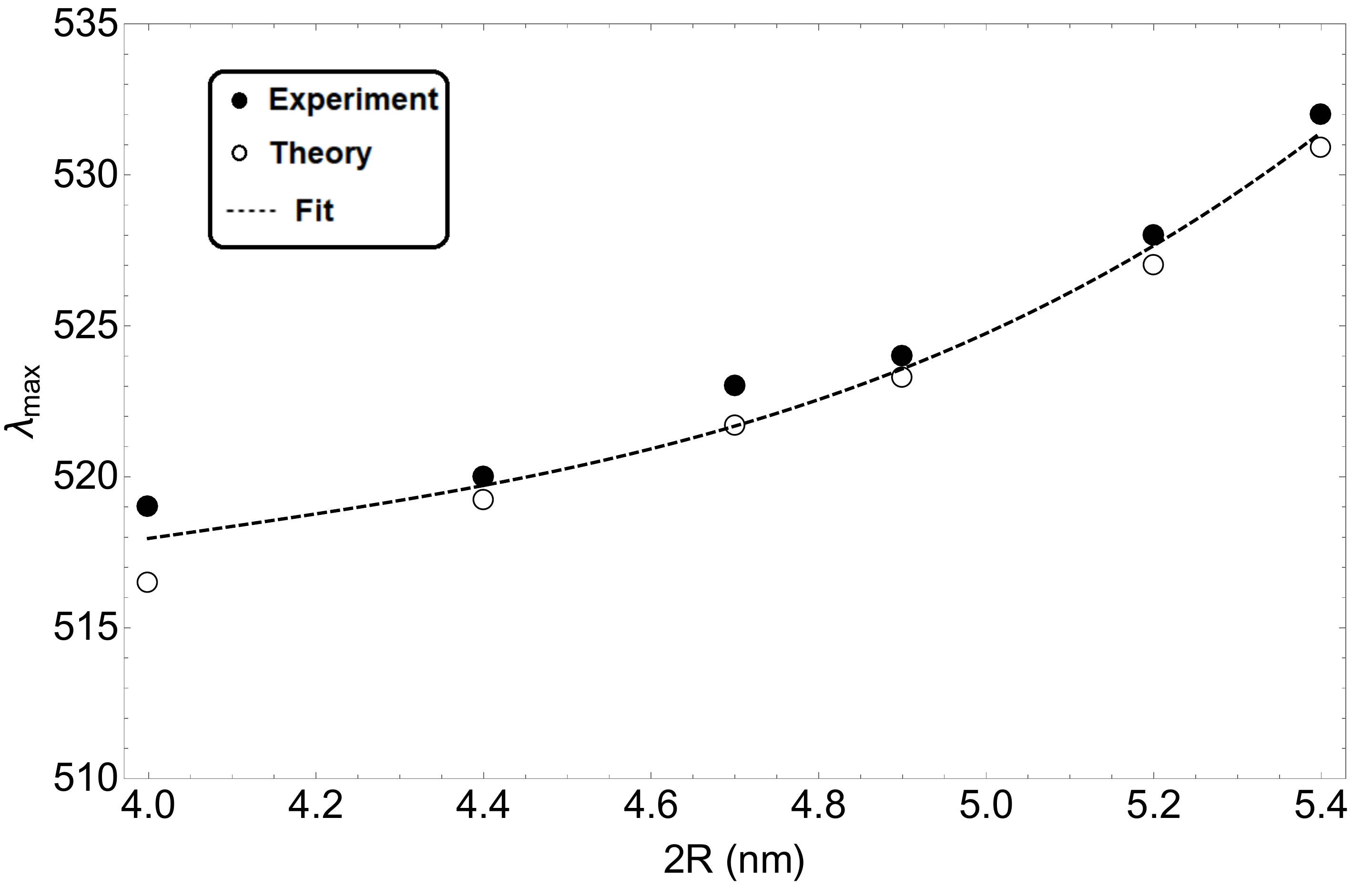}}
	\caption{Variations of surface plasmon resonance wavelength as a function of NP diameters. Dashed line in the fitted function.}
	\label{fig5}
\end{figure}

\section{Conclusions}
In a semi-classical phenomenological Drude-like model for determining the permittivity of an individual MNP, we proposed to consider restoration force term in the interaction of light with small NPs which can be called the classical confinement effect. For energy dissipation, we have considered all dominant damping mechanisms including electron-electron and electron-phonon scatterings, radiation and electron scattering by the NP surface.  In addition, in order to correct the shortages of theoretical background of dissipation mechanisms and take into account quantum confinement effect as well, we have taken into account a correction term to the damping factor obtained by the well-known theoretical studies. Numerical analysis is done for small gold NPs and the free parameters of system are determined by studying the existing experimental data related to the extinction cross section. Results show the good agreement between experiments and our model. Dynamical parameters obtained by this model can be very useful for future theoretical studies about the interaction of electromagnetic fields with MNPs in the linear and nonlinear optics of media containing such NPs and plasmonics field as well. 

\bibliography{refs}
\end{document}